\documentclass[floats,twocolumn,prb]{revtex4}
\usepackage{graphicx}
\begin{document}
\title{Disorder-Driven Superconductor-Insulator Transition in
$d$-Wave Superconducting Ultrathin Films}
\author{Long He}
\author{Jian Sun}
\author{Yun Song}
\email{yunsong@bnu.edu.cn} \affiliation{Department of Physics,
Beijing Normal University, Beijing 100875, China}
\date{\today}

\begin{abstract}
We study the superconductor-insulator transition (SIT) in $d$-wave
superconducting ultrathin films. By means of the kernel polynomial
method, the Bogoliubov-de Gennes equations are solved for square
lattices with up to $360\times 360$ unit cells self-consistently,
making it possible to observe fully the nanoscale spatial
fluctuations of the superconducting order parameters and
discriminate accurately the localized quasiparticle states from the
extended ones by the lattice-size scaling of the generalized inverse
participation ratio. It is shown that Anderson localization can not
entirely inhibit the occurrence of the local superconductivity in
strongly-disordered $d$-wave superconductors. Separated by an
insulating 'sea' completely, a few isolated superconducting
'islands' with significant enhancement of the local superconducting
order parameters can survive across the SIT. The disorder-driven
SIT, therefore, is a transition from a $d$-wave superconductor to a
Bose insulator which consists of localized Cooper pairs. Unlike an
$s$-wave superconductor which presents a robust single-particle gap
across the SIT, the optical conductivity of a $d$-wave
superconductor reveals a gapless insulating phase, where the SIT can
be detected by observing the disappearance of the Drude weight with
the increasing disorder.
\end{abstract}



\maketitle
\section{Introduction}
Recently, as huge amount of interest has been shown in the nanoscale
superconductivity of thin films and interfaces between complex
oxides, accompanied by the technological developments in the
synthesis of high-quality nanostructures, people are more concerned
about whether superconductivity can be enhanced in nanostructures
with respect to the bulk limit. \cite{Reyren,Franke,Bianconi}
Because disorder is generated by reducing the film thickness,
\cite{Schneider} we should take much account of the effect of
disorder in superconducting ultrathin films and heterointerfaces.
The effect of disorder in two-dimensional (2D) superconductors have
long been a subject of great interest. Because the 2D CuO$_2$ planes
are characteristic of all the high-temperature cuprate
superconductors, the analysis of disorder-induced inhomogeneities
can aid in the understanding of the nature and origin of the
$d$-wave superconducting state. \cite{Alloul, Balatsky} At atomic
limit, the cuprate superconductors can also be considered to have
heterostuctures with stacks of nanoscale superconducting planes
intercalated by charge reservoir layers with nanoscale periodicity.
\cite{Bianconi}

In 2D $d$-wave superconductors, the quasiparticle has a linear
dispersion relation in the vicinity of nodal points. Accordingly,
the elastic scattering of electrons caused by impurities is likely
to destroy Cooper pairs and suppress strongly the order parameter in
the vicinity of impurities. \cite{Millis}
For this reason, the pseudogap state of cuprates is predicted to
contain superconducting islands embedded in a normal metallic
matrix, resulting from the scattering effects of the
inhomogeneously-distributed pair breakers. \cite{Kresin,
Ovchinnikov} Some theoretical studies have found that the pairing
correlation of a disordered $d$-wave superconductor can be
significantly enhanced near the impurities, leading to a strong
spatial fluctuation of the coupling constant. \cite{Romer,
Khaliullin, Maska, Andersen, Nunner} Besides, the low-energy
quasiparticle states in disordered $d$-wave superconductors are
found to be localized with the presence of a mobility
gap,\cite{Franz96, Atkinson1} bringing about a
superconductor-insulator transition (SIT) at a critical disorder
strength when the pair-breaking effect of disorder is taken into
account fully. \cite{Scalettar}

A number of recent experiments \cite{Kopnov, Sacepe1, Lin, Sacepe2,
Baturina, Dubi} demonstrate a direct disorder-driven transition
between the superconducting and insulating phases in
highly-disordered thin films. It has also been discovered that a
superconducting state can be created in LaAlO$_3$/SrTiO$_3$
interface, \cite{Reyren,Caviglia} accompanied with an increase of
the relative disorder strength \cite{Bell} or the magnetic field
\cite{Mehta} across the SIT. Some theoretical researches return to
the problem of the SIT in 2D $s$-wave superconductors,
\cite{Feigelman1, Burmistrov, Bouadim, Feigelman2, Pokrovsky,
Altshuler, Feigelman3} but there is still debate about the nature of
the insulating phase and the mechanism that drives the SIT.
Recently, the insulating phase is predicted to be a Bose insulator,
represented by localized Cooper pairs with a nonzero pairing
amplitude. \cite{Fisher}

To address the spatial inhomogeneity of high-T$_c$ superconducting
thin films, it is increasingly important to have numerical methods
that are capable of simulating both microscopic and mesoscopic
length scales simultaneously. Beside, the inhomogeneity of the
pair-breaking effect should be accounted for in a fully
self-consistent manner. \cite{Franz97} The exact diagonalization
approach is one of the commonly adopted methods since it treats
precisely the disorder-induced scattering and presents entirely the
eigenvalues and eigenstates of a finite system. In addition, the
correct low-energy density of states (DOS) of inhomogeneous $d$-wave
superconductor produced by using the exact method cannot be obtained
when the self-consistent $T$-matrix approximation is taken, although
the correlation between impurity location and order parameter
variation has been preserved. \cite{Atkinson2, Atkinson3} However,
the exact calculation on large lattice is prevented primarily by the
memory limitation of computer. Besides, the self-consistency of the
Bogoliubov-de Gennes (BdG) equations \cite{Gennes} and the average
over a large number of disorder configurations demand the fast
processor speed strongly. Recently, the kernel polynomial method
(KPM) \cite{Weisse} is regarded as a distinct method for the
disordered systems because it allows for the numerical calculations
for dimensions of the order of $D\approx 10^9$. Therefore, the KPM
approach has the potential to calculate lattices two or three orders
of magnitude larger of the number of sites than the lattices studied
by the exact diagonalization method. Consequently, the
mesoscopic-scale inhomogeneity in 2D lattice can be fully presented
by the KPM calculations, and the accuracy can also be significantly
improved when the lattice-size scaling of a certain quantity is
performed for the extrapolation of results of finite lattices to the
infinite limitation.

In the present work, we apply the KPM approach to study the effect
of disorder in high-$T_c$ superconductors. For the large enough
lattices, the disordered $d$-wave superconductors are investigated
by observing rigorously the evolutions of the spatial fluctuations
of the superconducting order parameters with increasing disorder in
nanoscale. Since the KPM is designed to calculate the local density
of states (LDOS) instead of the eigenfunctions, we introduce the
generalized inverse participation ratio (GIPR) \cite{Song,Murphy} to
study the Anderson localization \cite{Anderson, Abrahams, Thouless}
of Bogoliubov quasiparticles.  The GIPR, which relies only on the
LDOS as defined in Eq.~(\ref{G2w}), has been proved to be a good
measure of Anderson localization. We find that all Bogoliubov
quasiparticles can be localized by weak disorder in the $d$-wave
ultrathin films. Analyzing the spatial fluctuations of the order
parameters, we show that several local superconducting islands can
survive across the SIT, suggesting that the insulating phase is
characterized by a Bose insulator. In addition, the carrier
localization can be manifested by the disappearance of the Drude
peak of the optical conductivity with increasing disorder. The
calculated results agree with the observed experimental facts.
\cite{Basov, Lobo, Uykur}

The paper is organized as follows: In Sec.~\ref{Model}, we describe
the KPM approach and the tight-binding BdG equations of the
disordered $d$-wave superconductors in detail; results are presented
in Sec.~\ref{Result-dSC}. We first extend the GIPR to confirm the
localization effect of disorder in $d$-wave supperconductors; then
we try to show a clear picture of the disorder-driven transition
from a $d$-wave superconductor to a Bose insulator; finally, we
discuss the experimental observation of the SIT in $d$-wave
superconductor by the optical conductivity, where the SIT is
accompanied by the disappearance of the Drude peak with the
increasing disorder. A concluding summary is given in
Sec.~\ref{conclusion}.

\section{Model and Methodology}
\label{Model}

We introduce a 2D mean-field Hamiltonian for the disordered $d$-wave
superconducting films, which is expressed as
\begin{eqnarray}
H&=&-t\sum_{\langle ij\rangle\sigma}c^{\dag}_{i\sigma}c_{j\sigma}
+\sum_{i\sigma}\epsilon_i c^{\dag}_{i\sigma} c_{i\sigma}\nonumber\\
&&-\sum_{\langle ij\rangle}\{\Delta_{ij} c_{i\uparrow}^{\dag}
c_{j\downarrow}^{\dag}+\textmd{H.c.}\}, \label{Ham}
\end{eqnarray}
where $c_{i\sigma}$ ($c_{i\sigma}^{\dag}$) are the electronic
annihilation (creation) operators at sites $i$ with spin $\sigma$
($\uparrow$ or $\downarrow$), $t$ denote the hopping integrals
between nearest neighbor (NN) sites $i$ and $j$, $\epsilon_i$
represent the on-site disorder energies distributed evenly in the
energy region [-W/2, W/2], and $\Delta_{ij}=-V\langle
c_{j\downarrow}c_{i\uparrow}\rangle$ are the superconducting order
parameters under the NN attractive interactions $V$ ($V<0$).

To take into account the $d$-wave gap symmetry in high-temperature
superconductors, we constraint
$\Delta_{ij}=\frac{1}{2}|\Delta_{ij}|[(-1)^{x_{ij}}-(-1)^{y_{ij}}]$
with ($x_{ij}$, $y_{ij}$) connecting sites $i$ and $j$. Using the
Bogoliubov transformation,\cite{Bogoliubov}  the eigenvalues
$E_{\alpha}$ of Hamiltonian Eq.~(\ref{Ham}) are obtained by solving
the BdG equations self-consistently, which are given by
\begin{eqnarray}
\sum_j
  \left(
     \begin{array}{cc}
        \xi_{ij} & \Delta_{ij}\\
        \Delta_{ij}^* & -\xi_{ij}
     \end{array}
  \right)
  \left(
     \begin{array}{cc}
        u_{\alpha}(r_j)\\
        v_{\alpha}(r_j)
     \end{array}
  \right)
  =E_{\alpha}
  \left(
     \begin{array}{cc}
        u_{\alpha}(r_i)\\
        v_{\alpha}(r_i)
     \end{array}
  \right), \label{BdGs}
\end{eqnarray}
with $\xi_{ij}=-t_{ij}+\epsilon_i\delta_{ij}$. Here $u(r_i)$ and
$v(r_i)$ denote the Bogoliubov coefficients of sites $i$. For a
lattice with $N$ sites, the Hamiltonian of BdG equations can be
represented by a $2N\times 2N$ Hermite matrix with $2N$ eigenvalues,
requiring $E_{N+k}=-E_k$ ($k=1,2,...,N$).

It is convenience to introduce two real-space Green's functions
$G^{11}(\omega)$ and $G^{12}(\omega)$ for Hamiltonian
Eq.~(\ref{Ham}), which are denoted by $N\times N$
matrices~\cite{Nagai}
\begin{eqnarray}
G^{11}_{ij}(\omega)&=&\sum_{l=1}^{N} \left(
\frac{u_{il}u^{*}_{jl}}{\omega-E_l}
+\frac{v_{il}v^{*}_{jl}}{\omega+E_l} \right),
\nonumber\\
G^{12}_{ij}(\omega)&=&\sum_{l=1}^{N} \left(
\frac{u_{il}v^{*}_{jl}}{\omega-E_l}
+\frac{v_{il}u^{*}_{jl}}{\omega+E_l} \right).\label{GFs}
\end{eqnarray}
As a result, LDOS with respect to sites $i$ can be expressed as
\begin{eqnarray}
\rho(\textrm{r}_i,\omega)&=&-\frac{1}{\pi}
\textrm{Im}G^{11}_{ii}(\omega)\nonumber\\
&=&\sum_{l=1}^N\{|u_{il}|^2\delta(\omega-E_l)
+|v_{il}|^2\delta(\omega+E_l) \}.
\end{eqnarray}
And similarly, the NN superconducting order parameters $\Delta_{ij}$
can be obtained by
\begin{eqnarray}
\Delta_{ij}&=&-\frac{V}{\pi}\int_{-\infty}^{+\infty}
\textrm{Im}G^{12}_{ij}(\omega)f(\omega)d\omega\nonumber\\
&=&-V\sum_{l=1}^N\{u_{il}v^{*}_{jl}f(E_l)
+v_{il}u^{*}_{jl}(1-f(E_l)) \},
\end{eqnarray}
where $f(E)$ represents the Fermi-Dirac distribution function.

Instead of directly diagonalizing the Hamiltonian matrix of BdG
equations,  the physical quantities, such as LDOS and
superconducting order parameters are calculated by expanding the
single particle Green's functions in terms of Chebyshev polynomials
within the KPM approach. \cite{Weisse,Weisse1} Considering that both
the first and second kinds of Chebyshev polynomials should be
defined in the interval [-1, 1], a simple linear transformation
should be introduced to rescale the Hamiltonian and all energies as
\begin{equation}
\widetilde{H}=\frac{H-b}{a},
~~\widetilde{\omega}=\frac{\omega-b}{a},
\end{equation}
where $a=(E_{max}-E_{min})/2$, and $b=(E_{max}+E_{min})/2$.
$E_{max}$ and $E_{min}$ are the extremal eigenvalues of Hamiltonian,
which can be approximately obtained on a small lattice by using the
Lanczos algorithm \cite{Dagotto}. Thus we can expand the single
particle Green's functions shown in Eq.~(\ref{GFs}) into series of
Chebyshev polynomials as
\begin{eqnarray}
\textrm{Im}\bar{G}_{ij}^{\alpha\beta}(\widetilde{\omega})&=&
-\frac{1}{\sqrt{1-\widetilde{\omega}^2}} \left(
g_0\mu_0^{\alpha\beta}
+2\sum_{n=1}^{M-1}g_n\mu_n^{\alpha\beta}T_n(\widetilde{\omega})
\right),\nonumber\\
\textrm{Re}\bar{G}_{ij}^{\alpha\beta}(\widetilde{\omega})&=&
-\frac{1}{\sqrt{1-\widetilde{\omega}^2}} \left(
g_0\mu_0^{\alpha\beta}
+2\sum_{n=1}^{M-1}g_n\mu_n^{\alpha\beta}U_n(\widetilde{\omega})
\right), \nonumber\\
\label{GF_KPM}
\end{eqnarray}
where $\alpha\beta=11$ or $12$,
$g_n=\sinh[\lambda(1-n/M)]/\sinh(\lambda)$  are the Lorenz Kernel
with a free parameter $\lambda$,  $M$ is the order of the series,
$\mu_n$ represent the coefficients of the expansions, and
$T_n(x)=\cos[n\arccos(x)]$ and
$U_n(x)=\sin[(n+1)\arcsin(x)]/\sin[\arccos(x)]$ are the Chebyshev
polynomials of the first and second kind respectively.

Now the pivotal issue has turned to how to calculate the
coefficients $\mu_n$ of the expansions. According to the proposal
given by Covaci {\it et al.},\cite{Covaci} the moments $\mu_n^{11}$
and $\mu_n^{12}$ can be obtained efficiently through a recursive
procedure. Since the Chebyshev polynomials have recursion relations
\begin{eqnarray}
&&T_0(\widetilde{H})=1, ~~~~~~~T_{-1}(\widetilde{H})=
T_{1}(\widetilde{H})=\widetilde{H},\nonumber\\
&&T_{m+1}(\widetilde{H})= 2\widetilde{H} T_{m}(\widetilde{H})-
T_{m-1}(\widetilde{H}),
\end{eqnarray}
we can obtain the coefficients by calculating the matric elements of
the Chebyshev polynomial as
\begin{eqnarray}
\mu_n^{11}(i,j)&=&\langle \textmd{vac}|c_{i\uparrow}T_n(\tilde{H})
c_{j\uparrow}^{\dag}|\textmd{vac}\rangle, \nonumber\\
\mu_n^{12}(i,j)&=&\langle
\textmd{vac}|c_{i\downarrow}^{\dag}T_n(\tilde{H})
c_{j\uparrow}^{\dag}|\textmd{vac}\rangle,
\end{eqnarray}
where $|\textmd{vac}\rangle$ is the vacuum state of Hamiltonian
Eq.~(\ref{Ham}). As a result, the whole Green's function is
extracted in absence of obtaining all the eigenvalues and
eigenstates.

\begin{figure}[h]
\includegraphics[width=20pc]{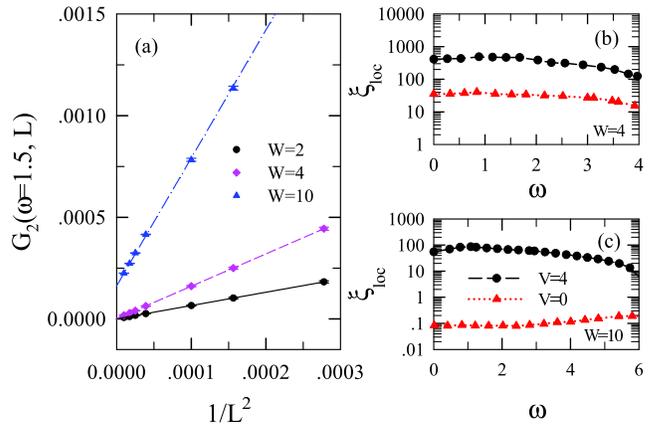}
\caption{(Color online) (a) The lattice size scaling of the
generalized inverse participation ratio $G_2(\omega, L)$ of a
quasiparticle with energy $\omega=1.5$ for attractive interactions
$V=4$, at different disorder strengths $W=2$, $W=4$, and $W=10$. The
localization lengths for $V=0$, and $V=-4$, at different disorder
strengths $W=4.0$ (b), and $W=10$ (c). The parameter of Lorenz
kernel is determined by $\lambda=\frac{M\cdot W}{L^2}$, and energies
are in unit of $t$.} \label{GIPR}
\end{figure}

\section{Disorder-Driven SIT in $d$-Wave Superconductors}
\label{Result-dSC}

The one-parameter scaling theory \cite{Abrahams} predicts that, in a
simple 2D Anderson model, arbitrarily weak disorder would localize
all electronic states. But a 2D superconductor is believed to
transition into an insulator beyond a certain critical value of the
disorder strength. There remains a dispute about whether the
insulating phase is characterized by a Bose insulator or a Fermi
insulator. \cite{Mehta} Here, we investigate the disorder-driven SIT
in $d$-wave superconductors from two aspects, i.e. the localization
of Bogoliubov quasiparticles and the Cooper pairs breaking across
the SIT.

\subsection{The Localization of Bogoliubov quasiparticles}

The localization length of a particle can be obtained by the GIPR,
which is defined as \cite{Song,Murphy}
\begin{equation}
G_2(\omega)=\frac{\sum_i \rho(\textrm{r}_i,\omega)^2} {[\sum_i
\rho(\textrm{r}_i,\omega)]^2},\label{G2w}
\end{equation}
where $\rho(\textrm{r}_i,\omega)$ represent the local density of
states (LDOS) at sites $i$. The GIPR is a proven way in testing the
Anderson localization of electronic states. \cite{Song,Murphy} Since
an energy broadening $\gamma$ has to be introduce to get a continuum
LDOS of a finite lattice,  a relation of $\gamma=W/L^{2}$ should be
satisfied to guarantee that we average over the same number of
states to calculate the GIPR for lattices of different sizes.
\cite{Murphy} For the Green's functions obtained by KPM, the
parameter $\lambda$ in Lorenz kernel plays a similar role as
parameter $\gamma$, except that the order $M$ of the Chebyshev
polynomial series also need to be taken in to account. Therefore, we
introduce the relation of $\lambda=\frac{WM}{L^2}$.

As shown in Fig.~\ref{GIPR}(a), the scaling of the GIPR satisfies a
linear relationship as
\begin{equation}
G_2(\omega, L)\doteq \alpha +\beta \frac{1}{L^2}, \label{LSGIPR}
\end{equation}
where $\alpha$ represents the intercept in the limitation
$L\rightarrow\infty$, and $\beta$ is the slope of the straight
scaling line. A localized state is predicted to have a nonzero
intercept, from which we can achieve the localization length by
\begin{equation}
\xi_{loc}(\omega)=\frac{1}{G_2(\omega,\infty)^{1/d}}=\frac{1}{\sqrt{\alpha}},
\end{equation}
where $d$ is the lattice dimension. It is also shown in
Fig.~\ref{GIPR}(a) that the intercept $\alpha$ increases
significantly with the strengthening of the disorder strength $W$,
demonstrating that the localization of the quasiparticles is
enhancement with the decreasing localization lengthes.

In Fig.~\ref{GIPR}(b) and \ref{GIPR}(c), we compare the localization
lengthes of Bogoliubov quasiparticles of 2D $d$-wave superconductors
with that of the electrons in 2D Anderson model, and find that the
localization lengthes increase by an order of magnitude or more when
the attractive interactions $V=4t$ are taken into account. Despite
this, the scaling of the GIPR predicts that disorder can still
localize all quasiparticles in 2D $d$-wave superconductors,
suggesting that disorder may introduce a direct transition from a
superconductor to an Anderson insulator. However, to discriminate a
Fermi insulator from a Bose insulator, the presence of both the
completely suppression to the superconductivity and the entirely
vanishing of the pair amplitude by disorder is essential. Next, we
study the effects of disorder on the superconducting order
parameters across SIT.

\begin{figure}[h]
\centering
\includegraphics[width=20pc]{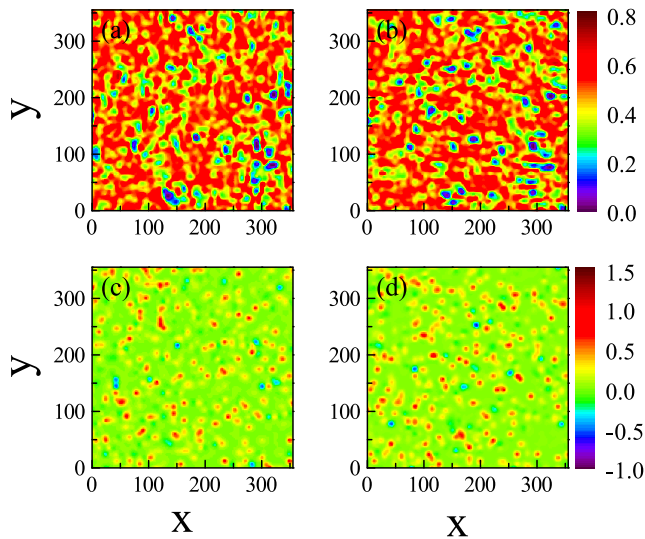}
\caption{(Color online) The disorder induced spatial fluctuations of
the superconducting order parameters $|\Delta_{i,i+\hat{x}}|$ at
different disorder strength $W=4$ (a) and $W=16$ (c). The
fluctuations of $|\Delta_{i,i+\hat{y}}|$ are also shown for $W=4$
(b) and $W=16$ (d), respectively. The other parameters are: L=360,
$N=L^2$, and $V=-4$. Energies are in unit of $t$.} \label{DSOP-W}
\end{figure}

\subsection{Superconducting Blobs}

It is well known that the suppression of superconductivity by
non-magnetic disorder is much stronger in the $d$-wave
superconductor than in the $s$-wave superconductor. Here we study
the competition and correlations between disorder and the $d$-wave
superconductivity in large-sized lattices. The NN superconducting
order parameters are acquired self-consistently by
\begin{equation}
\Delta_{ij}=-V\int^{E_c}_{-E_c}\textmd{Im}G^{12}_{ij}(E)
(1-2f(E))dE,
\end{equation}
where $G^{12}_{ij}$ is the off-diagonal Green's function, and $E_c$
(-$E_c$) represents the band-edge energy.  As expected, the average
of order parameters along one of the bond directions,
$\bar{\Delta}_{\eta}=1/N\sum_{i}\Delta_{i,i+\eta}$ with $\eta=
\hat{x}$ or $\hat{y}$, drops with the increasing disorder
monotonously. In spite of the maintenance of the $d$-wave symmetry
of the average order parameter of the whole lattice, the
superconductor is locally no longer $d$-wave because of the
violation of the translational invariance by disorder. In
Fig.~\ref{DSOP-W}, we plot the spatial distribution of the NN
superconducting order parameters $\Delta_{i,i+\hat{x}}$ and
$\Delta_{i,i+\hat{y}}$ in $d$-wave superconductors in weak ($W=4t$)
and strong ($W=16t$) disorder cases respectively.
Fig.~\ref{DSOP-W}(a) and \ref{DSOP-W}(b) show clearly that the
system is still a superconductor in the weak disorder cases.
However, the system transitions into an insulator as shown in
Fig.~\ref{DSOP-W}(c) and \ref{DSOP-W}(d) when $W=16t$, where the
great majority of the NN bonds are not superconducting at all but a
few local superconducting bonds can survive across the
disorder-driven SIT.

\begin{figure}[h]
\centering
\includegraphics[width=20pc]{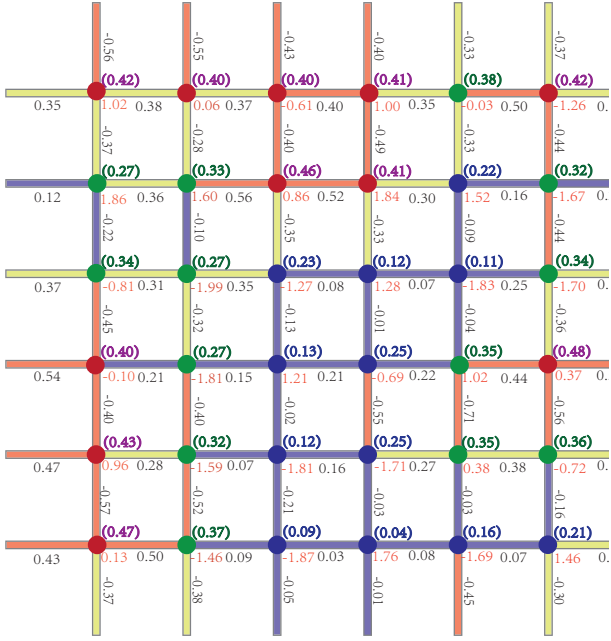}
\caption{(Color online) Real-space plot of NN superconducting order
parameters $\Delta_{ij}$ (numbers in black) and the on-site order
parameters $\Delta_i$ (numbers in brackets) of a certain part of a
square lattice with a particular realization of the random potential
$\epsilon_i$ denoted by the numbers in red. The bonds marked in red,
green and blue are corresponding to the NN order parameters with
$\Delta_i\geq 0.4$, $0.25<\Delta_i<0.4$, and $\Delta_i\leq 0.25$,
respectively. The colors of the filled circles are in the same rule
to display the compares values of the on-site order parameters
$\Delta_i$. The other parameters are: $V=-4$ and $W=4$. Energies are
in unit of $t$.} \label{SOP-bond}
\end{figure}

To understand whether the spatial fluctuations of
$\Delta_{i,i+\hat{x}}$ and $\Delta_{i,i+\hat{y}}$ are correlated or
independent, we illustrate in partial views of the effects of box
distributed disorder on the local superconducting order parameters
when the disorder strength is comparative with the attractive
interactions ($V=W=4t$). Concurrently, we also display the local
effects of disorder on the on-site $d$-wave superconducting order
parameters $\Delta_i$, which is expressed as
\begin{equation}
\Delta_i= \frac{1}{4}(\Delta_{i,i+\hat{x}} -\Delta_{i,i+\hat{y}}
+\Delta_{i,i-\hat{x}} -\Delta_{i,i-\hat{y}}).
\end{equation}
As shown in Fig.~\ref{SOP-bond}, it is obvious that the regions of
weakened and strengthen superconductivity are separated completely,
suggesting that the spatial fluctuations of the superconducting
order parameters along the two directions ($\Delta_{i,i+\hat{x}}$
and $\Delta_{i,i+\hat{y}}$) are not independent. Consequently, the
disorder energy of a certain site can not determine completely the
local on-site order parameters accordingly.

Because of the existence of correlation effects presented above, it
worth clarifying the relationship between the disorder energy and
the local superconducting order parameter. For the case without
disorder ($W=0$), the homogenous superconducting order parameters
are $\Delta_0=0.62t$ when $V=4t$. As the disorder is introduced,
there appears very strong spatial fluctuation of the superconducting
order parameters with the minimum value $|\Delta_{ij}^{min}|=0.01t$
and the maximum value $|\Delta_{ij}^{max}|=0.71t$ as shown in
Fig.~\ref{SOP-bond}. To define the system to be a Bose insulator, we
need to figure out why the superconducting order parameters can be
enhanced by disorder.

\begin{figure}[h]
\centering
\includegraphics[width=20pc]{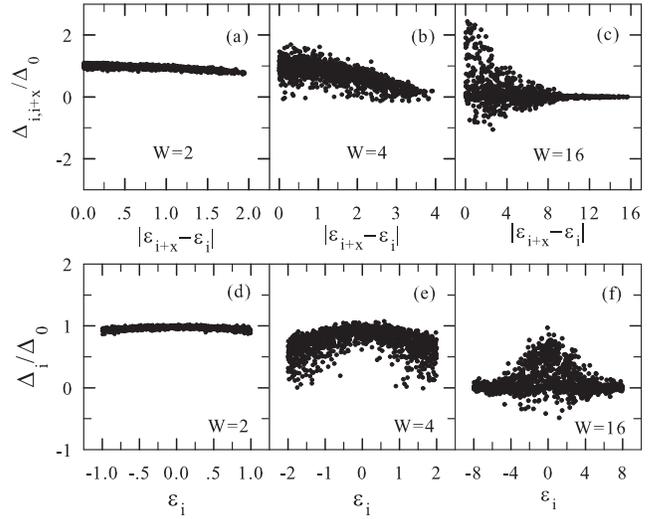}
\caption{Upper panel: the normalized NN superconducting order
parameters $\Delta_{i,i+x}/\Delta_0$ as a function  of the
energy-difference $|\epsilon_i-\epsilon_{i+x}|$ of the corresponding
sites at different disorder strengths $W=2$, $W=4$, and $W=16$.
$\Delta_0=0.62$ is the order parameter of a pure $d$-wave
superconductor with attractive interactions $V=-4$; Lower panel: the
normalized on-site order parameter $\Delta_i/\Delta_0$ as a function
of site energy $\epsilon_i$. The other parameters are: L=320,
$N=L^2$, and $V=-4$. Energies are in unit of $t$.} \label{SOPi}
\end{figure}

\subsection{A Bose insulator with localized Cooper pairs}

In Fig.~\ref{SOPi}, we plot the distributions of the NN order
parameters $\Delta_{i,i+\eta}$ (Fig.~\ref{SOPi}(a)-(c)) and the
on-site order parameters $\Delta_i$ (Fig.~\ref{SOPi}(d)-(f)) over
the disorder energies. It is shown that the lower the site energy
$\epsilon_i$ or the energy difference of bond
$|\epsilon_i-\epsilon_{i,i+\eta}|$ are, the stronger the suppression
of the superconducting order parameters can be observed. On the
contrary, the constrain effect of disorder is much weaker in the
lower-energy region, but the fluctuation of the order parameters is
getting stronger as shown in Fig.~\ref{SOPi}. To those sites with
zero disorder energy, the fluctuation is the strongest, suggesting
that the nonlocal effect of disorder can not be ignored. In the
strong disorder regime, we even find that the disorder potential can
change the sign of $\bar{\Delta}_{\eta}$ locally. As expected, the
fluctuations are getting stronger with the increasing of disorder
strength, and $\Delta^{max}_{ij}/\Delta_0$ increases to 2.43 when
$W=16t$ and $V=4t$. Therefore, separated completely by the very
large non-superconducting regions, there still exist several bonds
which have comparatively large superconducting order parameters.
Anderson's theorem \cite{Anderson} proposes that the transition
temperature and gap are insensitive to impurity scattering when the
coherent length $\xi_{coh}$ is much larger than the lattice spacing
$a$. Whereas we find strong fluctuations of the superconducting
order parameters in $d$-wave superconductor, suggesting that
Anderson's theorem is invalid. As a result, the coherent length
should be of the order of the lattice spacing $a$, implying that the
cooper pairs are strongly localized.

\begin{figure}[h]
\includegraphics[width=20pc]{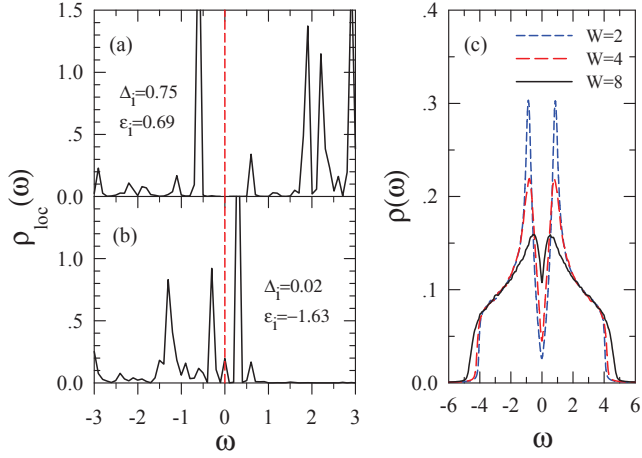}
\caption{(Color online) The local density of states at sites with
different local superconducting order parameters $\Delta_i=0.75$ (a)
and $\Delta_i=0.02$ (b) when $W=8$. (c) The density of states at
different disorder strength $W=2$, $W=4$, and $W=8$. The other
parameters are: $L=320$, $V=-4$, $M=2000$, and $\lambda=2$, Energies
are in unit of $t$.} \label{LDOS}
\end{figure}

It has been proposed that the presence of localized Cooper pairs
will lead to the disappearance of the coherence peaks in the
one-particle DOS whereas the superconducting gap remains intact.
\cite{Sacepe1} It is obvious that this judgement is currently only
applicable to the SIT driven by Cooper pair localization in some
disordered $s$-wave superconductors. \cite{Trivedi1} For the
$d$-wave superconductors, it has been demonstrated that even weak
disorder can significantly alter the DOS at low energy.
\cite{Atkinson1,Atkinson2} As a result, the DOS near the Fermi level
increase with the increasing disorder (Fig.~\ref{LDOS}(c)),
accompanied with the vanish of the coherence peaks of $d$-wave
Cooper pairs. In spite of this, a Bose insulator may be determined
by the detection of the localized Cooper pairs using energy
distribution analysis of the LDOS. As shown in Fig.~\ref{LDOS}(a)
and \ref{LDOS}(b), there is significant difference of the LDOS
around the Fermi surface between the lattice sites with or without
the localized Cooper pairs. Since the KPM approach allows us to do
calculations on large enough lattices, it becomes possible to
compare the theoretical and experimental results directly in
nanoscale, including the nanoscale inhomogeneities of the spatial
excitations and the LDOS features provided by the atomic-scale STM
experiments.

\subsection{SIT presented by optical Conductivity}

\begin{figure}[h]
\includegraphics[width=20pc]{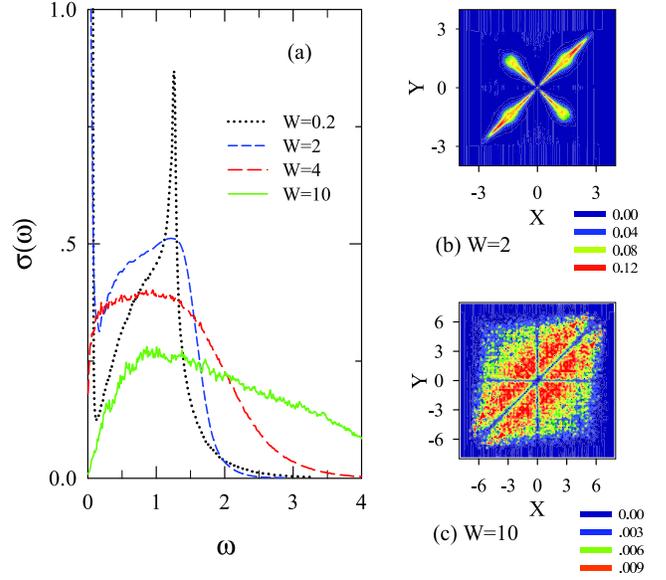}
\caption{(Color online) (a) The effect of disorder on the optical
conductivity. Insert: the corresponding DOS; (b) and (c) the matrix
element density $J(x,y)$ for the pure ($W=0$) and disordered ($W=4$)
systems, respectively. The other parameters are: $L=360$, $V=-4$,
$M=2000$, Energies are in unit of $t$.} \label{OCJxy}
\end{figure}

Next we study the disorder effect on the optical conductivity, which
can be calculated directly by \cite{Weisse,Weisse1},
\begin{eqnarray}
\sigma(\omega)&=&\sum_{k,q}\frac{| \langle
k|J|q\rangle|^2[f(E_k)-f(E_q)]} {2ZL^2\omega}\delta (\omega
-(E_k-E_q))
\nonumber\\
&=&\frac{1}{\omega}\int _{-\infty}^\infty
j(x,x+\omega)[f(x)-f(x+\omega)]dx,\label{OC-J}
\end{eqnarray}
where $Z$ is coordination number of the system. The matrix element
density $j(x,y)$ can be expanded as
\begin{equation}
j(x,y)=\sum^{M-1}_{n,m=0} \frac{\mu_{nm}h_{nm}g_ng_mT_n(x)T_m(y)}
{\pi^2\sqrt{(1-x^2)(1-y^2)}}
\end{equation}
where $g_n=1/\left(1+\delta_{0,n}\right)$ is the kernel damping
factors, $h_{nm}$ accounts for the correct normalization, and the
moments $\mu_{nm}$ are obtained from
\begin{eqnarray}
\mu_{nm}=
\int_{-1}^1\int_{-1}^1\tilde{j}(x,y) T_n(x)T_m(y)dxdy\nonumber\\
=\frac{1}{D}Tr\left[T_n(\tilde{H})JT_n(\tilde{H})J\right],
\label{mu}
\end{eqnarray}
where $\tilde{j}(x,y)$, which is transformed linearly from the
matrix element density $j(x,y)$, is defined within the energy
interval [-1, 1]. \cite{Weisse}

The disorder effects on the optical conductivity and matrix element
density $j(x,y)$ are plotted in Fig.~\ref{OCJxy}, respectively. We
find firstly that the optical conductance diverges near
$\omega=2\Delta_0=1.24t$ in the weakly disordered $d$-wave
superconductor. Meanwhile, the matrix element density is
concentrated in four small regions located symmetrically on the two
diagonal lines $x=\pm y$ with the so called "shark fan" structure as
shown in Fig.~\ref{OCJxy}(b). Eq.~(\ref{OC-J}) presents that the
Drude weight is obtained by calculating the integral of the density
of $j(x,y)$ along the diagonal line $x=y$. Since there exist the
low-energy quasiparticle excitations in the pure $d$-wave
superconductor, we find two peaks with "shark fin" structure along
the diagonal line $x=y$, corresponding to a significant Drude
weight. When the effects of disorder is considered, the the matrix
element density appears to be dispersing towards the regions
surrounded as shown in Fig.~\ref{OCJxy}(c). As a result, the Drude
weight tends to zero since the matrix element density along the
diagonal line $x=y$ drops very quickly with the increasing of
disorder strength. When the disorder strength is large than the
critical value $W_c\approx5t$ for the attractive interactions
$V=-4t$, we find that the spreading of the density of $j(x,y)$ in
the energy plan have six centers. As a result, no density of
$j(x,y)$ can stay in the diagonal line $x=y$, leading to the
completely suppression of the Drude weight by disorder in the high
disorder region. Therefore, it is convenient to probe the SIT in
$d$-wave superconductors by observing the Drude weight in the
optical conductivity.


\section{Conclusions}
\label{conclusion}

In summary, we extend the kernel polynomial method to investigate
the superconductor-insulator transition in inhomogenous two
dimensional $d$-wave superconductors. We have improved on previous
numerical results by calculating very big cluster, as well as come
up with explanations to the effects of disorder on nanoscale
superconductivity of thin films and interfaces between complex
oxides. It is manifest by the optical conductivity that disorder can
drive a transition from a superconductor to a gapless insulator in
the $d$-wave superconducting ultrathin films. The insulating phase
is characterized as a Bose insulator, where all Bogoliubov
quasiparticles are localized but a few of superconducting blobs can
survive strong disorder.

\section*{Acknowledgments}
We are very grateful to W. A. Atkinson for his valuable comments to
our manuscript. The computational resources utilized in this
research were provided by Shanghai Supercomputer Center. The work
was supported by the NSFC of China, under Grant No. 10974018 and
11174036, and the National Basic Research Program of China (Grant
Nos. 2011CBA00108).

\end{document}